\documentclass[aps,amsfonts,reprint,tightenlines,amssymb,superscriptaddress,twocolumn,floatfix]{revtex4-2}  

\usepackage{graphicx}% Include figure files
\usepackage{dcolumn}% Align table columns on decimal point
\usepackage{bm}% bold math
\usepackage{mathtools}
\usepackage{amsmath}
\usepackage{amsthm}

\usepackage{qcircuit}
\usepackage[caption=false]{subfig}

\usepackage{algorithm}
\usepackage[noend]{algpseudocode}
\algrenewcommand\algorithmicdo{}

\makeatletter
\renewcommand{\ALG@name}{Procedure}
\makeatother

\usepackage[linktocpage=true,
  colorlinks=true, 
  pdfborder={0 0 0},
  linkcolor=blue,
  citecolor=red,
  filecolor=yellow,
  urlcolor=blue,
  bookmarks,
  pdfauthor={},
]{hyperref}

\usepackage{ifthen}
\newcounter{is_qcircuit_used}
\begin{document}

\preprint{APS/123-QED}

\title{First-quantized adiabatic time evolution for the ground state of a many-electron system and the optimal nuclear configuration}

\author{Yusuke Nishiya}
\email{nishiya.y.aa@m.titech.ac.jp}
\affiliation{
Laboratory for Materials and Structures,
Institute of Innovative Research,
Tokyo Institute of Technology,
Yokohama 226-8503,
Japan
}
\affiliation{
Quemix Inc.,
Taiyo Life Nihombashi Building,
2-11-2,
Nihombashi Chuo-ku, 
Tokyo 103-0027,
Japan
}

\author{Hirofumi Nishi}
\affiliation{
Laboratory for Materials and Structures,
Institute of Innovative Research,
Tokyo Institute of Technology,
Yokohama 226-8503,
Japan
}
\affiliation{
Quemix Inc.,
Taiyo Life Nihombashi Building,
2-11-2,
Nihombashi Chuo-ku, 
Tokyo 103-0027,
Japan
}

\author{Yannick Couzinié}
\affiliation{
Laboratory for Materials and Structures,
Institute of Innovative Research,
Tokyo Institute of Technology,
Yokohama 226-8503,
Japan
}
\affiliation{
Quemix Inc.,
Taiyo Life Nihombashi Building,
2-11-2,
Nihombashi Chuo-ku, 
Tokyo 103-0027,
Japan
}

\author{Taichi Kosugi}
\affiliation{
Laboratory for Materials and Structures,
Institute of Innovative Research,
Tokyo Institute of Technology,
Yokohama 226-8503,
Japan
}

\affiliation{
Quemix Inc.,
Taiyo Life Nihombashi Building,
2-11-2,
Nihombashi Chuo-ku, 
Tokyo 103-0027,
Japan
}

\author{Yu-ichiro Matsushita}
\affiliation{
Laboratory for Materials and Structures,
Institute of Innovative Research,
Tokyo Institute of Technology,
Yokohama 226-8503,
Japan
}
\affiliation{
Quemix Inc.,
Taiyo Life Nihombashi Building,
2-11-2,
Nihombashi Chuo-ku, 
Tokyo 103-0027,
Japan
}
\affiliation{
Quantum Material and Applications Research Center,
National Institutes for Quantum Science and Technology (QST),
2-12-1, Ookayama, Meguro-ku, Tokyo 152-8552, Japan
}

\date{\today}

\begin{abstract}
We propose a novel adiabatic time evolution (ATE) method for obtaining the ground state of a quantum many-electron system on a quantum circuit based on first quantization.
%We propose an adiabatic time evolution (ATE) method for obtaining the ground state of a quantum many-electron system on a quantum circuit based on first quantization for the first time.
As a striking feature of the ATE method, it consists of only unitary operations representing real-time evolution, which means that it does not require any ancillary qubits, nor controlled real-time evolution operators.
Especially, we explored the first-quantized formalism of ATE method in this study, since the implementation of first-quantized real-time evolution on quantum circuits is known to be efficient.
However, when realizing the ATE quantum circuit in first-quantization formalism, obstacles are how to set the adiabatic Hamiltonian and how to prepare the corresponding initial ground state. 
We provide a way to prepare an antisymmetrized and non-degenerate initial ground state that is suitable as an input to an ATE circuit, which allows our ATE method to be applied to systems with any number of electrons.
In addition, by considering a first-quantized Hamiltonian for quantum-mechanical electron system and classical nuclear system, we design a quantum circuit for optimal structure search based on ATE.
Numerical simulations are demonstrated for simple systems, and it is confirmed that the ground state of the electronic system and optimal structure can be obtained by our method.
\end{abstract}

\maketitle 

\section{Introduction}
Efficient calculation for the ground state of a given Hamiltonian is of crucial importance in a wide range of fields.
This is because solving practically interesting problems can often be paraphrased as finding the ground state of a properly defined Hamiltonian.
In quantum physics and quantum chemistry, to prepare a ground state and low-energy states is also important because a lot of fundamental properties can be extracted from these states of the electron system.
Research into quantum algorithms for ground-state preparation has been actively pursued \cite{Kadowaki1998PhysRevE, Farhi2000arXiv, Aspuru-Guzik2005Science, Poulin2009PhysRevLett, Ge2019JMathPhys, Lin2020nearoptimalground, Lin2022PRXQ, Choi2021PRL, Silva2021arXiv, Kosugi2022PhysRevResearch, kosugi2022exhaustive, Meister2022arXiv, Stetcu2022arXiv, Xie2022arXiv, Chan2023arXiv}.

Several schemes have been proposed to obtain the ground state by realizing non-unitary operations to the system of interest on quantum circuits, such as imaginary-time evolution (ITE)~\cite{Silva2021arXiv, Kosugi2022PhysRevResearch, kosugi2022exhaustive, Xie2022arXiv, Chan2023arXiv}, cosine functions~\cite{Choi2021PRL, Meister2022arXiv}, or shifted step functions~\cite{Lin2020nearoptimalground, Lin2022PRXQ}.
What these methods have in common is that they introduce ancillary qubits to widen the Hilbert space, exploit the probabilistic nature in the observations of ancillary qubits, and utilize forward and backward controlled real-time evolution (CRTE) operators.

On the other hand, there are widely known methods for ground-state calculation called adiabatic quantum computation (AQC) or quantum annealing (QA)~\cite{Kadowaki1998PhysRevE, APOLLONI1989233, Albash2018RevModPhys}, and these have attracted attention in the field of combinatorial optimization~\cite{Lucas2014, Lechner2015} as well as quantum chemistry~\cite{xia:2018aa,copenhaver:2021aa,teplukhin:2020aa,teplukhin:2021aa,teplukhin:2022aa}.
Recently, QA has also been gaining attention in the field of materials science, particularly in terms of predicting crystal structures~\cite{Gusev2023Nature, couzinie2023annealing} and adsorption sites of molecules~\cite{Sampei2023JACS_Au}. 
In these methods, energy minimization problems for atoms and molecules interacting under some modeled potential are translated into quadratic unconstrained binary optimization (QUBO) or higher-order unconstrained binary optimization (HUBO) form, which is compatible with the various Ising machines currently available.
As examples of the application of AQC to quantum chemistry, some schemes for ground-state preparation have been proposed~\cite{Babbush2014srep, Aspuru-Guzik2005Science, VeisPittner2014JChemPhys, Shirakawa2021PhysRevResearch, Sugisaki2022CommunChem}, all of which are based on second quantization.
By QA based on the first-quantized Hamiltonian and using the kinetic term as a quantum fluctuation,
Koh and Nishimori~\cite{YWKoh2022PhysRevA} discuss the optimization problem of Shinomoto-Kabashima-like potential, which has multiple local minima in a continuous one-dimensional space.
However, there is no mention of how to construct a quantum circuit representing the real-time evolution (RTE) by the first-quantized Hamiltonian.

%Kassal et al.~\cite{Kassal2008PNAS} proposed a quantum algorithm for computing the RTE of a quantum system based on the first-quantized Hamiltonian and demonstrated that quantum computers can simulate the RTE in polynomial time while the computational cost using classical computers increases exponentially with system size.
The RTE of a quantum system based on the first-quantized Hamiltonian on a quantum computer was proposed in Ref.~\cite{Lloyd1996Science, Lloyd1997PhysRevLett}, and Kassal et al.~\cite{Kassal2008PNAS} demonstrated that quantum computers can simulate the RTE in polynomial time while the computational cost using classical computers increases exponentially with system size.
Moreover, the advantage of employing first quantization over second quantization is discussed in Ref.~\cite{Jones2012NJP, Kosugi2022PhysRevResearch, kosugi2022exhaustive}.
The scaling of the operation number per RTE step with respect to the number of electrons $n_e$ is estimated to be better when considering the first-quantized Hamiltonian than the second-quantized one.
Specifically, it is evaluated as $\mathcal{O}(n_e^2 \mathrm{poly}(\mathrm{log}n_e))$ for the first quantization while it is $\mathcal{O}(n_e^4)$ for the second quantization due to the two-electron integrals in the Hamiltonian~\cite{Helgaker}.
In this paper, we describe a method to obtain the ground state of a many-electron system using first-quantized adiabatic time evolutioin (ATE) on a quantum circuit and give an example of the construction of an appropriate initial Hamiltonian and its ground state.
%As already mentioned, since their ITE implementation is transformed into CRTE implementation in their scheme, the advantage of the first-quantization translates equivalently to other RTE based quantum algorithms and thus in particular to our ATE scheme. 
ATE does not require ancillary qubits, observations in the middle of calculation, nor CRTE operations because of its unitarity.

In addition, by considering a first-quantized Hamiltonian for quantum-mechanical electron system and classical nuclear system, we construct a quantum circuit for optimal structure search on the Born-Oppenheimer surface, using ATE in a similar way as in Ref.~\cite{kosugi2022exhaustive}.
Hirai et al.~\cite{Hirai2022ACSOmega} presented a molecular structure optimization method where nuclei are treated as quantum mechanical particles, but in general more qubits are required to represent the nuclear wavefunction.
In contrast to the method of structural optimization by calculating energy gradients proposed in Ref.~\cite{Kassal2009JChemPhys}, our structural optimization scheme does not require multiple observations of the gradients and furthermore is a global search in the possible coordination space.
We have applied these schemes to simple systems and performed numerical simulations.

\section{Methods}
\subsection{Implementation of adiabatic time evolution}
\subsubsection {Adiabatic quantum computation}
Let us consider the time-dependent Hamiltonian $\hat{\mathcal{H}}(t)$ which evolves from $\hat{\mathcal{H}}( 0)=\hat{\mathcal{H}}_\mathrm{ini}$ to $\hat{\mathcal{H}}(t_\mathrm{f})=\hat{\mathcal{H}}_\mathrm{fin}$.
%\begin{gather}
%    \hat{\mathcal{H}}(t) = A_1(t) \hat{\mathcal{H}}_\mathrm{f}+\left(1-A_2(t)\right) \hat{\mathcal{H}}_\mathrm{i}
%\end{gather}
%where the boundary conditions for scheduling functions are
%$A_i(0)=0$ and $A_i(t_\mathrm{f})=1$ for all $i \in \mathbb{N}^+ \equiv \{ 1,2,3, \cdots \}$. 
$\hat{\mathcal{H}}_\mathrm{ini}$ and $\hat{\mathcal{H}}_\mathrm{fin}$ represent the initial and the final Hamiltonian, respectively. 
$|\psi_t^\mathrm{gs}\rangle$ is the instantaneous ground state of $\hat{\mathcal{H}}(t)$ when considering the time-independent Schrödinger equation $\hat{\mathcal{H}}(t)|\psi \rangle = \epsilon_t |\psi \rangle$. By discretizing time into $N$ steps with the constant interval $\Delta t$, the time-evolved state from $t=0$ to $t=t_\mathrm{f}$ according to the time-dependent Schrödinger equation can be approximated within the first-order Trotter splitting as
\begin{gather}
    |\psi(t_\mathrm{f})\rangle \approx \prod_{m=N}^1 \left[ e^{ -i \hat{\mathcal{H}}(t_m) \Delta t } \right] |\psi(0)\rangle,
    \label{def_psi_tf}
\end{gather}
with $\Delta t \equiv t_\mathrm{f} / N$ and $t_m \equiv m \Delta t$. 
$\Delta t$ is assumed to be sufficiently small.
If $|\psi(0)\rangle = |\psi_0^\mathrm{gs}\rangle$ is satisfied , $t_\mathrm{f}$ is large, and the change of $\hat{\mathcal{H}}(t)$ is very slow, $|\psi(t_\mathrm{f})\rangle \approx |\psi_{t_\mathrm{f}}^\mathrm{gs}\rangle$ is derived~\cite{Ehrenfest, BornFock}.
The ground state of $\hat{\mathcal{H}}_\mathrm{fin}$ results from an ATE starting from the well-known Hamiltonian $\hat{\mathcal{H}}_\mathrm{ini}$ and its ground state. The following is conventionally used as the condition for the time evolution to be adiabatic~\cite{Kato1950, Amin2009PhysRevLett}:
\begin{gather}
    \max_{j \in \{ 1,2,3, \cdots \}, t \in[0, t_\mathrm{f}]} \left \{ \frac{|\langle \psi_t^j |\partial_t \hat{\mathcal{H}}(t) |\psi_t^\mathrm{gs}\rangle |}{\Delta_j (t) ^2} \right \}\ll 1,
    \label{adiabatic_condition}
\end{gather}
where $|\psi_t^j \rangle$ is the $j$-th excited state of the instantaneous $\hat{\mathcal{H}}(t)$ and $\Delta_j(t)$ is the energy gap between the instantaneous ground state and the $j$-th excited state. 
The transitions from the ground state at $t$ are assumed to be dominant since $| \psi(0)\rangle$ is initially $|\psi_0^\mathrm{gs}\rangle$.
\subsubsection{First-quantized Hamiltonian}
Here we target a first-quantized Hamiltonian for an electronic system, which is expressed as the sum of the kinetic energy part $\hat{T}$ and the position-dependent potential part $V_\mathrm{fin}$ or $V_\mathrm{ini}$:
\begin{gather}
    \begin{dcases}
    \hat{\mathcal{H}}_\mathrm{fin} = \hat{T} + V_\mathrm{fin}(\{ \hat{\boldsymbol{r}}_l \}), \\
    \hat{\mathcal{H}}_\mathrm{ini} = \hat{T} + V_\mathrm{ini}(\{ \hat{\boldsymbol{r}}_l \}) ,
    \end{dcases}
\end{gather}
where
\begin{gather}
    \hat{T} \equiv \sum_{l}\frac{\hat{\boldsymbol{p}}_l^2}{2m_e}.
\end{gather}
$\hat{\boldsymbol{p}}_l$ and  $\hat{\boldsymbol{r}}_l$ are the momentum and the position operators of the $l$-th electron, respectively. 
$m_e$ is the electron mass. 
Both $V_\mathrm{fin}$ and $V_\mathrm{ini}$ consist of external potentials and electron-electron interactions.
All the quantities in this paper are in atomic units unless otherwise stated. 
We define the time-dependent Hamiltonian for ATE as
\begin{gather}
    \hat{\mathcal{H}}(t)=\hat{T} + \hat{V}(t),
    %\hat{V}(t) \equiv A_1(t)V_\mathrm{f}(\{ \hat{\boldsymbol{r}}_l \}) + (1-A_2(t))V_\mathrm{i}(\{ \hat{\boldsymbol{r}}_l \})
\end{gather}
where the boundary conditions of the potential part are $\hat{V}(0)= V_\mathrm{ini}$ and $\hat{V}(t_\mathrm{f})= V_\mathrm{fin}$.
It is natural to assume that $\hat{V}(t)$ for every $t$ is symmetric with respect to the exchange of any two electrons since they are identical particles.
By employing a first-order Suzuki-Trotter expansion in conjunction with the adiabatic theorem, the ground state $|\psi_{t_\mathrm{f}}^\mathrm{gs} \rangle$ of the objective Hamiltonian $\hat{\mathcal{H}}_\mathrm{fin}$ is given as
\begin{gather}
    |\psi_{t_\mathrm{f}}^\mathrm{gs} \rangle \approx \prod_{m=N}^1 \left [ e^{ -i \hat{T} \Delta t }
    \cdot e^{-i \hat{V}(t_m) \Delta t } \right ] |\psi_0^\mathrm{gs} \rangle .
\end{gather}

In this study, the initial Hamiltonian $\hat{\mathcal{H}}_\mathrm{ini}$ is chosen so that its ground state is not degenerate for simplicity although the adiabatic condition in the case where the ground state is allowed to be degenerate is discussed in Ref.~\cite{Jansen2007JMathPhys}.
%If the ground state  of the initial Hamiltonian is $n$-fold degenerate, what is obtained by ATE is a superposition from the ground state to the $(n-1)$-th excited state of the final Hamiltonian.
%Furthermore, the initial ground state $|\psi(0)\rangle = |\psi_0^{gs} \rangle $ must be antisymmetric with respect to the exchange of two electrons. 
%Otherwise, since the $\hat{\mathcal{H}}(t)$ is symmetrically constructed with respect to the exchange of two electrons, Eq. (\ref{def_psi_tf}) shows that the final state $|\psi(t_\mathrm{f}) \rangle$ is not antisymmetric and physically meaningless.
Furthermore, we have to keep in mind that the final state $|\psi(t_\mathrm{f}) \rangle$ must be antisymmetric with respect to the exchange of any pair of electrons~\cite{Abrams1997PRL}.
Otherwise the final state obtained has no physical meaning.
Now, since the time-dependent Hamiltonian $\hat{\mathcal{H}}(t)$ is assumed to be symmetric with respect to the exchange of any pair of electrons,
it follows immediately from Eq. (\ref{def_psi_tf}) that the final state is antisymmetric if and only if the initial ground state $|\psi(0)\rangle = |\psi_0^\mathrm{gs} \rangle $ is antisymmetric.
An example of how to construct a non-degenerate and antisymmetrized initial ground state for a many-electron system is described in Sec. \ref{sec:initial_state_preparation}.

\subsubsection{Construction of the quantum circuit}
\label{circuit_el}
We encode the $n_e$-electron wavefunction confined in a cubic cell of size $L$ using $n_{qe}$ qubits for each spatial direction per electron, as usual in the first-quantized formalism~\cite{Wiesner1996arxiv, Zalka1998PRSLA, Kassal2008PNAS, Kosugi2022PhysRevResearch, Jones2012NJP, Chan2023sciadv}. 
The spin degree of freedom for each electron can be encoded by introducing a single additional qubit per electron. We ignore, however, the spin degrees of freedom of spins in the present study for simplicity.
We refer to the $3n_e n_{qe}$ qubits collectively as the electronic register.
We generate  uniform grid points in the cell to encode the wavefunction $\psi$ as
\begin{gather}
    |\psi\rangle=\Delta V^{n_e / 2} \sum_{\boldsymbol{k}_0, \ldots, \boldsymbol{k}_{n_e-1}} \psi\left(\boldsymbol{r}^{\left(\boldsymbol{k}_0\right)}, \ldots, \boldsymbol{r}^{\left(\boldsymbol{k}_{n_e-1}\right)}\right) \nonumber \\
    \cdot\left|\boldsymbol{k}_0\right\rangle_{3 n_{q e}} \otimes \cdots \otimes\left|\boldsymbol{k}_{n_e-1}\right\rangle_{3 n_{q e}},
\end{gather}
where $\boldsymbol{k}_l$ is the vector of the three integers specifying the position eigenvalue $(k_{l_x}\boldsymbol{e}_x + k_{l_y}\boldsymbol{e}_y + k_{l_z}\boldsymbol{e}_z)\Delta x$ for the $l$-th electron. 
$\Delta x \equiv L / N_{q e}$ is the grid spacing of $N_{qe} \equiv 2^{n_{qe}}$ grid points in each spatial direction.
$\Delta V \equiv \Delta x^3 $ is the volume element for the normalization of $|\psi \rangle$.

\begin{figure*}[ht]
        \centering
        \includegraphics[width=0.7 \textwidth]{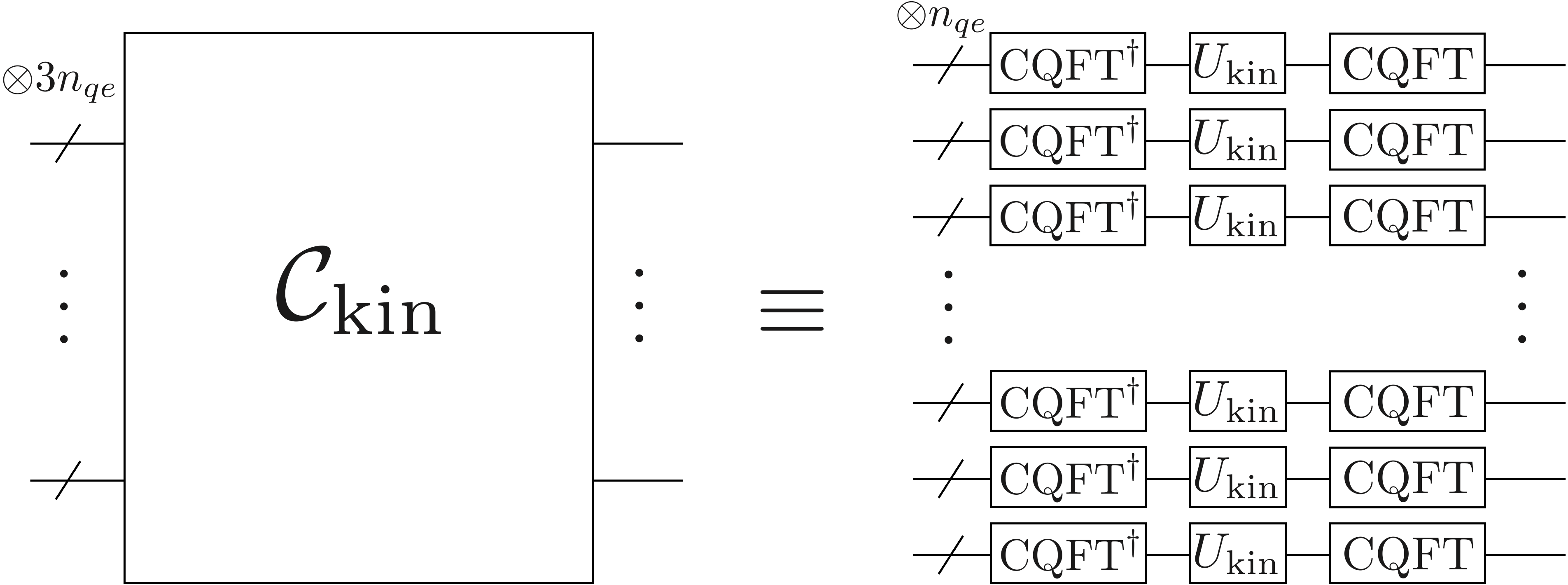}
 \caption{
    Quantum circuit of $\mathcal{C}_\mathrm{kin}$ which implements the RTE due to $\hat{T}$ for the $n_e$-electron system. 
    In the left circuit, the top wire bundle corresponds to the first electron and the bottom bundle corresponds to the $n_e$-th electron
 }
 \label{fig:Ckin}
\end{figure*}

As the canonical counterpart of the discretized positions, we define the $N_{qe}$ discrete momenta of a particle for each spatial direction $p^{(\widetilde{s})} \equiv \widetilde{s} \Delta p(\widetilde{s}=-N_{qe} / 2,-N_{qe} / 2+1, \cdots , N_{qe}/2-1)$ with the momentum step $\Delta p \equiv 2\pi / L$ in reciprocal space. 
The tilde symbol for an integer $j$ means
$\tilde{j} \equiv j - N_{qe}/2$ in what follows. We define the momentum eigenstate specified by three integers $s_x$, $s_y$, and $s_z$ as the
Fourier transform of the position eigenstates:
\begin{gather}
    \left|\boldsymbol{p}^{(\tilde{\boldsymbol{s}})}\right\rangle \equiv \frac{1}{N_{qe}^{3 / 2}} \sum_{k_x=0}^{N_{qe}-1} \sum_{k_y=0}^{N_{qe}-1} \sum_{k_z=0}^{N_{qe}-1} \exp \left(i \boldsymbol{p}^{(\tilde{\boldsymbol{s}})} \cdot \boldsymbol{r}^{(\boldsymbol{k})}\right)|\boldsymbol{k}\rangle_{3 n_{qe}}.
\end{gather}
$\left|\boldsymbol{p}^{(\tilde{\boldsymbol{s}})}\right\rangle$ is also the eigenstate of the kinetic energy operator $\hat{T}_\nu \equiv \hat{p}_\nu^2 / (2m)\; (\nu=x,y,z)$ for each direction belonging to the discrete kinetic energy $E_{\mathrm{kin} s} \equiv \widetilde{s}^2(\Delta p)^2 /(2 m)$.
By defining the kinetic-phase gate $U_\mathrm{kin}(\Delta t)$ for $n_{qe}$ qubits such that it acts on a computational basis $|j\rangle _{n_{qe}}$ as $U_\mathrm{kin}(\Delta t)|j\rangle _{n_{qe}} = \exp(-iE_{\mathrm{kin}j} \Delta t)|j\rangle _{n_{qe}}$, the $n_{qe}$-qubit real-time evolution (RTE) operator generated by $\hat{T}_\nu$ can be implemented by using the centered quantum Fourier transform (CQFT)~\cite{Somma2015arxiv, Ollitrault2020PRL} as 
\begin{gather}
    \mathrm{CQFT} \cdot U_{\mathrm{kin}}(\Delta t) \cdot \mathrm{CQFT}^{\dagger}=e^{-i \hat{T}_\nu \Delta t} .
\end{gather}
Therefore, the circuit $\mathcal{C}_\mathrm{kin}$, which represents $e^{-i\hat{T}\Delta t}$, the real-time evolution by the kinetic energy part of the $n_e$-electron system, can be constructed as shown in Fig. \ref{fig:Ckin}.

\begin{figure*}[ht]
        \centering
        \includegraphics[width=0.75 \textwidth]{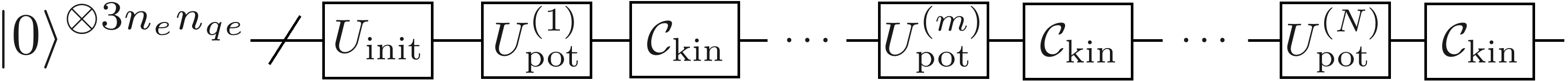}
 \caption{
    Quantum circuit to obtain the ground state of $n_e$-electron system by ATE based on first-order Suzuki-Trotter expansion. After the ground state of the initial Hamiltonian is created by $U_\mathrm{init}$, ATE operation discretized into $N$ steps follows.
 }
 \label{fig:C_ATE_el}
\end{figure*}

$U_\mathrm{pot}^{(m)}$ is the real-time evolution by the potential part at $t=t_m$:
\begin{gather}
    U_\mathrm{pot}^{(m)} \equiv e^{-i \hat{V}(t_m) \Delta t}.
\end{gather}
The entire circuit is shown in Fig. \ref{fig:C_ATE_el}.
$U_\mathrm{init}$ is an initialization gate to make the antisymmetrized ground state wavefunction of the initial Hamiltonian.

If $\hat{V}(t)$ consists of the sum of the two-body interactions and the one-body external potential terms for all $t$ as

\begin{gather}
    \hat{V}(t)= \sum_{l=1}^{n_e}v_\mathrm{ext}(\hat{\boldsymbol{r}}_l;t) + \sum_{l<l'}v(|\hat{\boldsymbol{r}}_l-\hat{\boldsymbol{r}}_{l'}|;t),
\end{gather}
then the same method as in Ref.~\cite{kosugi2022exhaustive} can be employed in order to implement $U_\mathrm{pot}^{(m)}$, and the circuit depth and the operation number of one ATE step $e^{-i\hat{\mathcal{H}}(t_m)\Delta t}$ is estimated as $\mathcal{O}(n_e^2 \mathrm{poly}(\log n_e))$, thanks to the employment of position eigenstates in the basis of each electron's register.
$v_\mathrm{ext}(\hat{\boldsymbol{r}}_l)$ and $v(|\hat{\boldsymbol{r}}_l-\hat{\boldsymbol{r}}_{l'}|)$ are assumed to be approximated by piecewise polynomials.

On the other hand, in the second quantization the Hamiltonian becomes
\begin{gather}
    \hat{\mathcal{H}} = \sum_{i,j} h_{ij}a_i^\dagger a_j +  \frac{1}{2}\sum_{i,j,k,l} h_{ijkl}a_i^\dagger a_j^\dagger a_k a_l,
    \label{second_q_H}
\end{gather}
where $a_i^\dagger$ and $a_i$ are fermionic creation and annihilation operators for the molecular orbitals, and the coefficients $h_{ij}$ and $h_{ijkl}$ are the one- and two-electron integrals, respectively.
The number of terms in the Hamiltonian is dominated by the second term in Eq.~\ref{second_q_H}, the two-body interaction part, and estimated to be $\mathcal{O}(n_e^4)$ under the assumption that the number of molecular orbitals to be considered is proportional to the number of electrons $n_e$.
Hence, the operation number per ATE step scales as $\mathcal{O}(n_e^4)$  in general~\cite{Babbush2015PhysRevA}.

\subsubsection{Preparation of the non-degenerate and antisymmetrized initial ground state of many-electron system}
\label{sec:initial_state_preparation}
In practice, it is quite difficult to know in advance how many dimensions the eigenspace of the ground state of the objective Hamiltonian has and how the instantaneous spectrum structure of $\hat{\mathcal{H}}(t)$ changes.
One approach to obtaining the ground state of the objective Hamiltonian as the output of ATE in such cases is to construct an initial Hamiltonian whose ground state is not degenerate.
It should be further noted that since we are encoding the wavefunction based on the first quantization, if we consider a many-electron system, the initial ground state must be antisymmetric with respect to the exchange of any pair of electrons.
However, the preparation of the initial ground state satisfying these conditions is nontrivial.
In this section, we describe an example of how to create a non-degenerate and antisymmetrized initial ground state of an $n_e$-electron system.

Here, we consider an initial Hamiltonian with no interactions between electrons:
\begin{gather}
    \hat{\mathcal{H}}_\mathrm{ini} = \sum_{l=1}^{n_e} \left[ \frac{\boldsymbol{\hat{p}}_l}{2m_e} + v_\mathrm{ini}(\hat{\boldsymbol{r}}_l)\right].
    \label{H_nele}
\end{gather}
The one-electron Hamiltonian $\hat{\mathcal{H}}_1$ is then defined as
\begin{gather}
    \hat{\mathcal{H}}_1 \equiv \frac{\boldsymbol{\hat{p}}}{2m_e} + v_\mathrm{ini}(\hat{\boldsymbol{r}}).
\end{gather}
Let the eigenvalues of $\hat{\mathcal{H}}_1$ be denoted by $\epsilon_1, \epsilon_2, \cdots$ in ascending order, and the corresponding eigenfunctions be $\phi_1(\boldsymbol{r}), \phi_2(\boldsymbol{r}), \cdots$.
These eigenfunctions are reffered to as one-electron orbitals.
The antisymmetrized many-electron wavefunction of the ground state of $\hat{\mathcal{H}}_\mathrm{ini}$ is represented by the Slater determinant formed by $n_e$ one-electron orbitals from $\phi_1$ to $\phi_{n_e}$:
\begin{gather}
    \Psi_\mathrm{i}(\boldsymbol{r}_1, \boldsymbol{r}_2, \cdots ,\boldsymbol{r}_{n_e}) 
    =
    \frac{1}{\sqrt{n_e !}}\left|\begin{array}{ccc}\phi_1\left(\boldsymbol{r}_1\right) & \cdots & \phi_1\left(\boldsymbol{r}_{n_e}\right) \\ \vdots & \ddots & \vdots \\ \phi_{n_e}\left(\boldsymbol{r}_1\right) & \cdots & \phi_{n_e}\left(\boldsymbol{r}_{n_e}\right)\end{array}\right|.
\end{gather}
Berry et al.~\cite{Berry2018npjQI} proposed an efficient technique to make a many-electron wavefunction represented by a single Slater determinant on a quantum circuit.
The ground state energy $E_\mathrm{i}^\mathrm{gs}$ of the $n_e$-electron system is
\begin{gather}
    E_\mathrm{i}^\mathrm{gs} = \sum_{l=1}^{n_e} \epsilon_l,
\end{gather}
where $E_\mathrm{i}^\mathrm{gs}$ is not degenerate when the highest occupied orbital and the lowest unoccupied orbital in the one-electron orbitals are not degenerate.
Therefore, whether the initial ground state is degenerate or not depends on $v_\mathrm{ini}$, the number of electrons $n_e$, and the dimension of space.
However, if $v_\mathrm{ini}$ is chosen such that the one-electron orbitals are not degenerate, then for any $n_e$, the ground state of the many-electron system constructed above is not degenerate.
This can be achieved, for example, by making $v_\mathrm{ini}$ an anisotropic harmonic potential:
\begin{gather}
 \hat{\mathcal{H}}_1 = \sum_{\mu = x,y,z} \left[ \frac{\hat{p}_\mu^2}{2m_e} + \frac{1}{2}m_e\omega_\mu^2\hat{r}_\mu^2 \right].
\end{gather}
As is well known, an eigenvalue of the above one-electron Hamiltonian are specified by a set of three integers $(n_x, n_y, n_z)$ which are greater than or equal to zero:
\begin{gather}
 \epsilon(n_x, n_y, n_z) = \sum_{\mu = x,y,z} \omega_\mu \left( n_\mu + \frac{1}{2} \right).
\end{gather}
If the angular frequencies are set as $\omega_x=1, \omega_y=\sqrt{2}$, and $\omega_z = \sqrt{3}$, we have the following statement that is proved in Appendix~\ref{sec:proof_of_LI} and implies that one-electron orbitals have no degeneracy:
\begin{gather}
    (n_x, n_y, n_z) \neq (m_x, m_y, m_z) \nonumber \\ 
    \Rightarrow  \epsilon(n_x, n_y, n_z) \neq \epsilon(m_x, m_y, m_z).
    \label{injection}
\end{gather}

For some prepared initial Hamiltonians, you can utilize the variational quantum circuit in Ref.~\cite{horiba2023construction} to generate the superposition of many Slater determinants.

\subsection{Search for the optimal structure of molecule}
\label{method_str_opt}
\subsubsection{Setup} 
Here we consider $n_e$ quantum mechanical electrons and $n_\mathrm{nucl}$ classical nuclei system as in the case of structural optimization of molecular systems using probabilistic imaginary-time evolution (PITE)~\cite{kosugi2022exhaustive}. Thus the objective Hamiltonian is
\begin{gather}
    \hat{\mathcal{H}}_\mathrm{fin} = \underbrace{\sum_{l=1}^{n_e} \frac{\hat{\boldsymbol{p}}_l^2}{2m_e}}_{\equiv \hat{T}_\mathrm{el}}
    + \underbrace{\sum_{l=1}^{n_e}v_\mathrm{ext}(\hat{\boldsymbol{r}}_l)}_{\equiv \hat{V}_\mathrm{ext}}
    + \underbrace{\frac{1}{2}\sum_{l\neq l'}v_{ll'}^\mathrm{(ee)}(|\hat{\boldsymbol{r}}_l-\hat{\boldsymbol{r}}_{l'}|)}_{\equiv \hat{V}_\mathrm{ee}} \nonumber \\
    +\underbrace{\sum_{l=1}^{n_e} \sum_{\lambda=1}^{n_\mathrm{nucl}} v_{l \lambda}^\mathrm{(en)}(|\hat{\boldsymbol{r}}_l-\boldsymbol{R}_{\lambda}|)}_{\equiv \hat{V}_\mathrm{en}} \nonumber \\
    + \underbrace{\frac{1}{2} \sum_{\lambda \neq \lambda '} v_{\lambda \lambda '}^\mathrm{(nn)}(|\boldsymbol{R}_{\lambda}-\boldsymbol{R}_{\lambda '}|)}_{\equiv \hat{V}_\mathrm{nn}},
    \label{final_hamiltonian}
\end{gather}
where $v_\mathrm{ext}$, $v^\mathrm{(ee)}$, $v^\mathrm{(en)}$, and $v^\mathrm{(nn)}$ represent the external potential for an electron, electron-electron interaction, electron-nucleus interaction, and nucleus-nucleus interaction, respectively. Note that the kinetic term of nuclei is ignored and the nuclear position $\boldsymbol{R}_\lambda$ appears as a classical parameter.
To find the ground state of this Hamiltonian, 
we consider a quantum register using a total of $3 n_e n_{qe} + n_{qn}$ qubits as
%we encode the following wavefunction on a quantum register using a total of $3 n_e n_{qe} + n_{qn}$ qubits.
\begin{gather}
    |\Psi \rangle = \sum_{\boldsymbol{J}} \sqrt{w_{\boldsymbol{J}}} |\psi [\boldsymbol{J}] \rangle_{3 n_e n_{qe}} \otimes |\boldsymbol{J}\rangle_{n_{qn}},
    \label{wf_stropt}
\end{gather}
% where $n_d$ is the dimension of space.
where $n_{qn}$ is the number of qubits allocated to a register $|\boldsymbol{J} \rangle_{n_{qn}}$ representing a possible nuclear configuration specified by the vector $\boldsymbol{J}$.
%$|\boldsymbol{J} \rangle_{n_{qn}}$ is an $n_{qn}$-qubit register that represents a possible nuclear configuration specified by the vector $\boldsymbol{J}$. 
Due to the quantum superposition, up to $2^{n_{qn}}$ different structures can be entered at once. $| \psi [\boldsymbol{J}] \rangle_{3 n_e n_{qe}}$ is an $3 n_e n_{qe}$-qubit register representing the many-electron wavefunction for each configuration $\boldsymbol{J}$ in three dimensional space. 
The overall quantum register $|\Psi \rangle$ can be written as a superposition of $|\psi [\boldsymbol{J}] \rangle_{3 n_e n_{qe}} \otimes |\boldsymbol{J}\rangle_{n_{qn}}$, and when $|\Psi_\mathrm{f}^\mathrm{gs} \rangle$ is obtained as the ground state of $\hat{\mathcal{H}}_\mathrm{fin}$, the most stable nuclear configuration can be determined by measuring the $n_{qn}$ qubits assigned to the nuclear configuration part.

The aim is to find the ground state of $\hat{\mathcal{H}}_\mathrm{fin}$ in Eq. (\ref{final_hamiltonian}) by ATE. We consider the time-dependent Hamiltonian
\begin{gather}
    \hat{\mathcal{H}}(t)=\hat{T}_{\mathrm{el}} \otimes \hat{I}_{\mathrm{nucl}}+A_1(t)\hat{V}_\mathrm{ext}\otimes \hat{I}_\mathrm{nucl} \nonumber \\
    + A_2(t) \hat{V}_{\mathrm{ee}} \otimes \hat{I}_{\mathrm{nucl}}+A_3(t) \hat{V}_{\mathrm{en}} \nonumber \\
    +A_4(t) \hat{I}_{\mathrm{el}} \otimes \hat{V}_{\mathrm{nn}} \nonumber \\
    +(1-A_5(t))\hat{V}_0 \otimes \hat{I}_{\mathrm{nucl}} \nonumber \\
    -\left(1-A_6(t)\right) \hat{I}_{\mathrm{el}} \otimes J_x  \sum_{l=1}^{n_{qn}} \hat{X}_l,
\end{gather}
where the boundary conditions for functions $A_i$ ($i\in\{ 1,2,3, \cdots \}$) are $A_i(0)=0$ and $A_i(t_\mathrm{f})=1$.
The initial and final Hamiltonians are thus given by
\begin{gather}
    \hat{\mathcal{H}}(0)= (\hat{T}_{\mathrm{el}} + \hat{V}_0 ) \otimes \hat{I}_{\mathrm{nucl}}- \hat{I}_{\mathrm{el}} \otimes J_x  \sum_{l=1}^{n_{qn}} \hat{X}_l,
    \label{initial_H_stropt} \\
    \hat{\mathcal{H}}(t_\mathrm{f})=\hat{\mathcal{H}}_\mathrm{fin}.
\end{gather}
$\hat{T}_\mathrm{el}$, $\hat{V}_\mathrm{ext}$, $\hat{V}_\mathrm{ee}$, $\hat{V}_0$, and $\hat{I}_\mathrm{el}$ act on the electronic wavefunction part denoted by $| \psi [\boldsymbol{J}] \rangle_{3n_e n_{qe}}$ in Eq. (\ref{wf_stropt}) and $\hat{I}_\mathrm{el}$ is the identity operator for the electron-part of the quantum register.  
$\hat{V}_\mathrm{nn}$,  $\hat{X}_l$, and $\hat{I}_\mathrm{nucl}$ act on the nuclear configuration part denoted by $|\boldsymbol{J}\rangle_{n_{qn}}$ and $\hat{I}_\mathrm{nucl}$ is the identity operator for the nuclear part. 
$\hat{V}_0$ is the initial potential of the electronic system and depends on the position operators of the electrons.
It is arbitrary as long as it is symmetric for the exchange of any pair of electrons and the ground state of $\hat{T}_\mathrm{el}+\hat{V}_0$ is available as the input for the quantum circuit described in Sec. \ref{circuit_opt}. 
$\hat{X}_l$ is defined as
\begin{gather}
    \hat{X}_l \equiv \underbrace{\hat{I}_2 \otimes \cdots \otimes \hat{I}_2}_{l-1\text{ qubits}} \otimes \underbrace{\hat{\sigma}_x}_{l\text{-th}} \otimes \underbrace{\hat{I}_2 \otimes \cdots \otimes \hat{I}_2}_{n_{qn}-l \text{ qubits}},
\end{gather}
where ${\hat{I}_2}$ and $\hat{\sigma}_x$ are the identity operator and the Pauli-X gate for a single qubit, respectively.

The second term of Eq. (\ref{initial_H_stropt}) is introduced inspired by the transverse field term in conventional AQC or QA, which allows the nuclear part of the initial ground state to be a uniform superposition of all nuclear configurations.
Since there is no interaction between the electronic part and the nuclear part in Eq. (\ref{initial_H_stropt}), the ground state $|\Psi_0^\mathrm{gs} \rangle$ of the initial Hamiltonian $\hat{\mathcal{H}}(0)$ can be straightforwardly constructed as the tensor product of the ground states of each part as
\begin{gather}
    |\Psi_0^\mathrm{gs} \rangle = |\psi_0^\mathrm{gs} \rangle_{3n_e n_{qe}} \otimes |+\rangle^{\otimes n_{qn}},
\end{gather}
where $|\psi_0^\mathrm{gs} \rangle_{3n_e n_{qe}}$ and $|+\rangle \equiv (|0\rangle + |1\rangle)/\sqrt{2}$ are the ground states of $\hat{T}_{\mathrm{el}} + \hat{V}_0$ and $-\hat{\sigma}_x$, respectively.

\subsubsection{Quantum circuit for structural optimization by ATE} 
\label{circuit_opt}
 We define the time-evolution operators at $t=t_m$ as
\begin{gather}
 \begin{dcases}
    U_\mathrm{ext}^{(m)} \equiv \exp \left[{-i\cdot A_1(t_m)\hat{V}_\mathrm{ext}\cdot \Delta t} \right],\\
    U_\mathrm{ee}^{(m)} \equiv \exp \left[{-i\cdot A_2(t_m)\hat{V}_\mathrm{ee}\cdot \Delta t} \right], \\
    U_\mathrm{en}^{(m)} \equiv \exp \left[{-i\cdot A_3(t_m)\hat{V}_\mathrm{en}\cdot \Delta t} \right], \\
    U_\mathrm{nn}^{(m)} \equiv \exp \left[{-i\cdot A_4(t_m)\hat{V}_\mathrm{nn}\cdot \Delta t} \right], \\
    U_0^{(m)} \equiv \exp \left[{-i\cdot (1-A_5(t_m)) \hat{V}_0 \cdot \Delta t} \right].
 \end{dcases}
\end{gather}
The time-evolution by the transverse-field-like term can be rewritten using $x$-axis rotation gates $R_x(\theta) \equiv e^{-i\frac{\theta}{2}\hat{\sigma}_x}$ as 
\begin{gather}
    [R_x(\theta_m)]^{\otimes n_{qn}} = \exp \left[ i \Delta t (1-A_6(t_m)) \cdot J_x \sum_{l=1}^{n_{qn}} \hat{X}_l \right],
    \label{eq:_RX_transverse}
\end{gather}
where the rotation angles are defined as $\theta_m \equiv -2\Delta t(1-A_6(t_m))J_x$.

\begin{figure*}[ht]
        \centering
        \includegraphics[width=0.75 \textwidth]{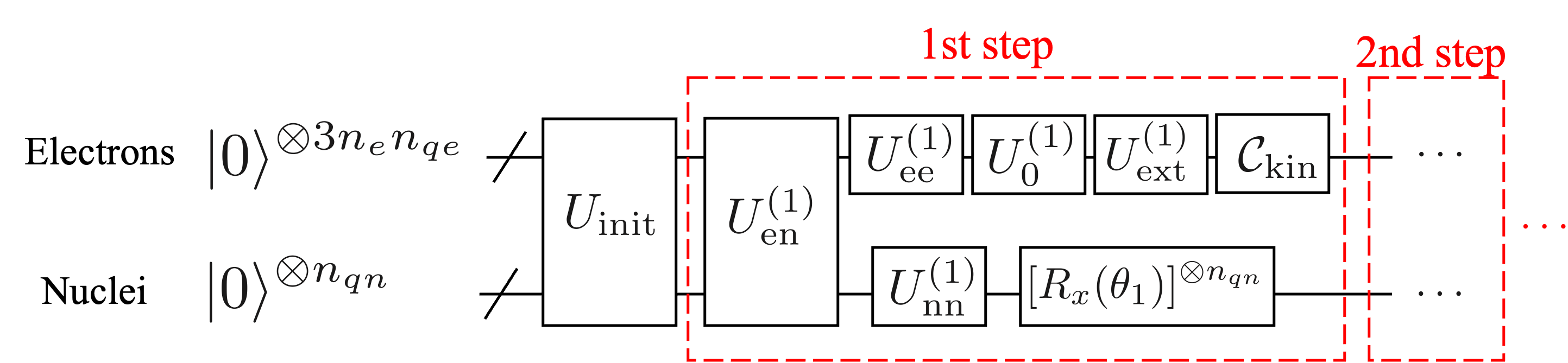}
 \caption{
    Quantum circuit to find the most stable nuclear configuration among up to $2^{n_{qn}}$ candidates. $n_e$-electron system in three-dimensional space are modelled and $3 n_{qe}$ qubits per electron are  used to represent the wavefunction of the electron system.
    After the ground state of the initial Hamiltonian is created by $U_\mathrm{init}$, ATE operation discretized into $N$ steps follows.
 }
 \label{fig:C_ATE_stropt}
\end{figure*}

As in the case of the electronic system only, we use time discretization and the first-order Suzuki-Trotter expansion to construct a quantum circuit for ATE.
The ground state $| \Psi_\mathrm{f}^\mathrm{gs}\rangle$ of $\hat{\mathcal{H}}(t_\mathrm{f})$ can be approximated as a result of ATE as
\begin{gather}
    |\Psi_\mathrm{f}^\mathrm{gs} \rangle \approx \nonumber \\
    \prod_{m=N}^1 \left [ \left ( e^{-i\hat{T}\Delta t} U_0^{(m)} U_\mathrm{ee}^{(m)} \otimes  [R_x(\theta_m)]^{\otimes n_{qn}} U_\mathrm{nn}^{(m)} \right ) \cdot U_\mathrm{en}^{(m)} \right ] \nonumber \\
    \cdot |\Psi_0^\mathrm{gs} \rangle.
\end{gather}
The schematic of the quantum circuit for the optimal structure search is depicted in Fig. \ref{fig:C_ATE_stropt}, using $\mathcal{C}_\mathrm{kin}$ described in Sec. \ref{circuit_el}.
The transverse field-derived part of Eq. (\ref{eq:_RX_transverse}) is the $x$-axis rotation for each qubit, hence the circuit depth is $\mathcal{O}(1)$.
If $\hat{V}_0$ consists of a two-body interaction term between electrons and a one-body external potential term,
the circuit depth per ATE step is typically estimated as $\mathcal{O}(n_e^2 \mathrm{poly}(\log n_e))$ according to the implementation of RTE in Ref.~\cite{kosugi2022exhaustive}.

\section{Applications}
\subsection{Definitions of scheduling functions and infidelity}
In this section, we consider a one-electron system in one-dimensional space of length $L$ as simple applications of our approach. We examine the following slightly simplified Hamiltonian:
\begin{gather}
    \hat{\mathcal{H}}(t) = \hat{T}+A(t) V(\hat{x}).
    \label{one_ele_H}
\end{gather}
The ground state of the initial Hamiltonian $\hat{\mathcal{H}}(0)=\hat{T}$ is a uniform superposition if $\hat{T}$ is implemented using CQFT. This can be easily prepared by initializing every qubit to the $|0\rangle$ state and then applying a Hadamard gate $H$ to every qubit:
\begin{gather}
    |\psi(0) \rangle = |\psi_0^\mathrm{gs} \rangle = |+ \rangle^{\otimes n_{qe}},\\
    U_\mathrm{init} = H^{\otimes n_{qe}}.
\end{gather}

From the adiabatic condition Eq. (\ref{adiabatic_condition}) with $\hat{\mathcal{H}}(t)$ in Eq. (\ref{one_ele_H}) , $t_\mathrm{f}$ should be set to satisfy the condition
\begin{gather}
    t_\mathrm{f} \gg \max_{j\in \{ 1,2,3,\cdots \}, s\in [0,1]}  \left \{  \frac{|\langle \psi_{st_\mathrm{f}}^j |V|\psi_{st_\mathrm{f}}^\mathrm{gs}\rangle |}{\Delta_j (st_\mathrm{f}) ^2} \cdot \frac{dA}{ds}(st_\mathrm{f}) \right \},
    \label{adiabatic_c_tf}
\end{gather}
where $s \equiv t/t_\mathrm{f}$ is normalized time.
Here we consider monotonous schedules $A$ and define the function $f(A)$ with the dimension of time as
\begin{gather}
    f(A(st_\mathrm{f})) \equiv \max_{j\in \{ 1,2,3, \cdots \} } \left \{ \frac{|\langle \psi_{st_\mathrm{f}}^j |V|\psi_{st_\mathrm{f}}^\mathrm{gs}\rangle |}{\Delta_j (st_\mathrm{f}) ^2} \right \}.
    \label{def_f}
\end{gather}
The right-hand side of  Eq. (\ref{def_f}) can be calculated from the eigenvectors and eigenvalues of $\hat{T}+A\cdot V$. 
Inserting this into Eq. (\ref{adiabatic_c_tf}) gives
\begin{gather}
     t_\mathrm{f} \gg \max_{s\in [0,1]} \left \{ f(A) \cdot \frac{dA}{ds}\right \}.
     \label{ac_w_f}
\end{gather}
Regarding the scheduling function $A$, which controls the strength of the potential at each time, we consider two types of functions, denoted as $A_\mathrm{lin}$ and $A_\mathrm{opt}$. They are constructed to satisfy the following equations
\begin{gather}
    \begin{dcases}
    A_\mathrm{lin}(t) = \frac{t}{t_\mathrm{f}}, \\
    \frac{dA_\mathrm{opt}}{ds}(st_\mathrm{f})=\frac{c}{f(A_\mathrm{opt}(st_\mathrm{f}))}.
    \end{dcases}
    \label{A_opt}
\end{gather}
The constant $c$ with the same dimension as time is determined by the boundary conditions $A_\mathrm{opt}(0)=0$ and $A_\mathrm{opt}(t_\mathrm{f})=1$.
$A_\mathrm{opt}$ is designed so that $dA_\mathrm{opt}/ds$ is small when $f(A_\mathrm{opt})$ is large, with the aim of making the right-hand side of Eq. (\ref{ac_w_f}) small.
Note that $A_\mathrm{opt}$ is ideal rather than practical since the spectral structure and the eigenvectors of the Hamiltonian at each time must be known in advance to solve Eq. (\ref{A_opt}). 
For these scheduling functions, the adiabatic condition Eq. (\ref{ac_w_f}) becomes as
\begin{gather}
    \begin{dcases}
    t_\mathrm{f} \gg \max_{A\in[0,1]} f(A) &\text{for $A_\mathrm{lin}$}, \\
    t_\mathrm{f} \gg c &\text{for $A_\mathrm{opt}$}.
    \label{ac_final}
    \end{dcases}
\end{gather}

We define the infidelity $\delta_N$ after ATE to $t_\mathrm{f}$ over $N$ steps as
\begin{gather}
    \delta_N= 1-| \langle \phi_0 | \psi (t_\mathrm{f}; N) \rangle |^2,
\end{gather}
where $| \phi_0 \rangle=|\psi^\mathrm{gs}_{t_\mathrm{f}}\rangle$ is the ground state of the final Hamiltonian $\hat{\mathcal{H}}_\mathrm{fin}$ obtained by numerical diagonalization, and $|\psi (t_\mathrm{f}; N) \rangle$ denotes the right-hand side of Eq. (\ref{def_psi_tf}) with $N$ explicitly written as a parameter.
In the following numerical simulations, the ATE circuits are configured as shown in Fig. \ref{fig:C_ATE_el} or Fig. \ref{fig:C_ATE_stropt}, and are based on the first-order Suzuki-Trotter expansion. All simulations are carried out using Qiskit, an open-source library for quantum circuit simulation~\cite{Qiskit}.

\subsection {One electron in a parabolic potential}
\label{sec_harm_ele}
As the first example, we calculated the ground state of a single electron in a parabolic potential
\begin{gather}
    V(\hat{x})=\frac{1}{2}m_e \omega^2\left(\hat{x}-\frac{L}{2}\right)^2,
\end{gather}
where the parameters are set as $\omega=1$ and $L=10$. In the simulation, we set $\Delta t=0.1$ and $n_{qe}=6$, i.e.\ we discretize the space into $2^6=64$ grid points.

\begin{figure}[ht]
    \centering
    \includegraphics[width=0.45 \textwidth]{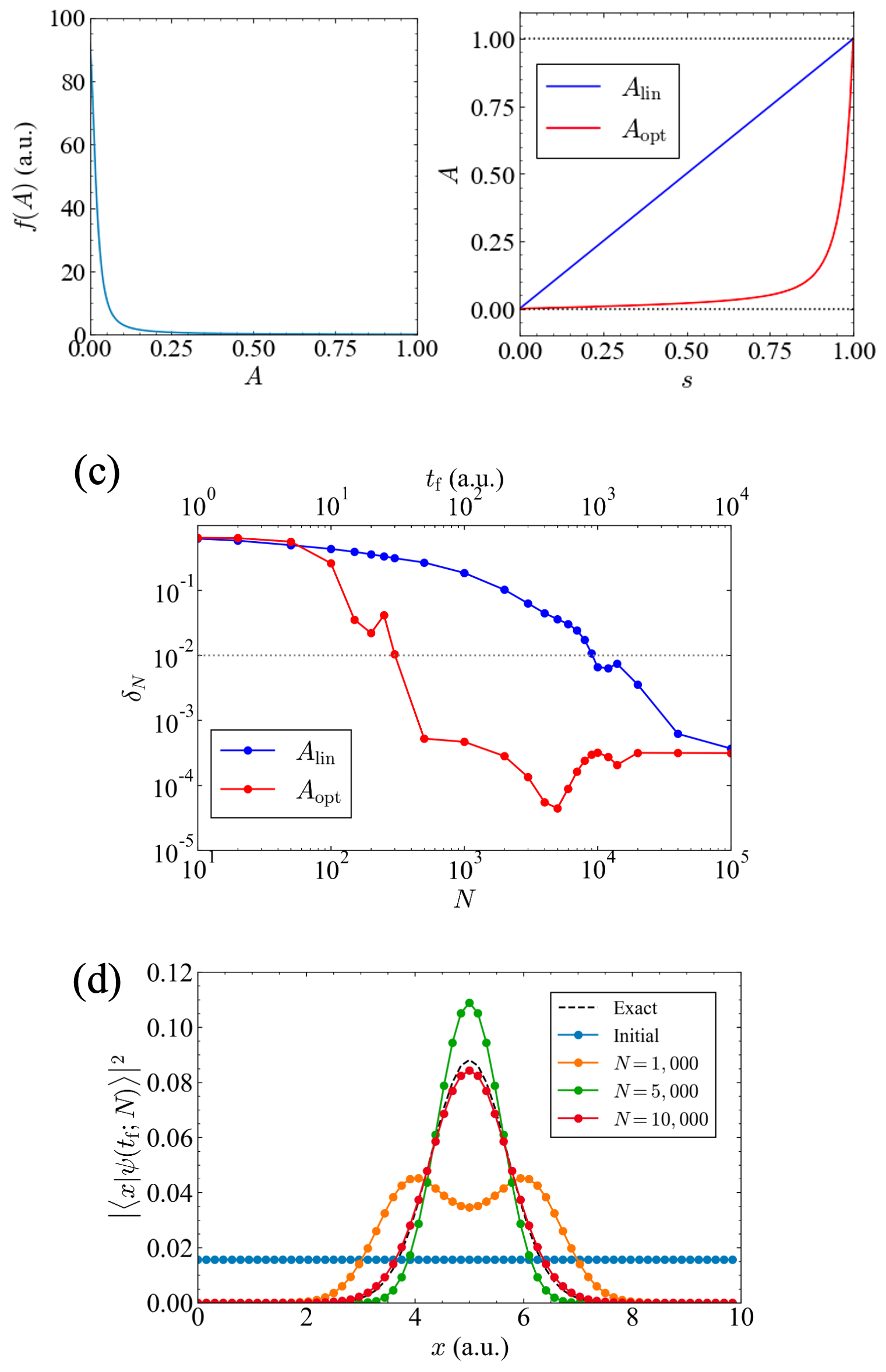}
    \caption{
    The simulation results of ATE for one electron system under the parabolic potential.
    (a) The indicator of adiabatic condition $f(A)$ defined in Eq. (\ref{def_f}). The maximum value is 92.01
    (b) The two scheduling functions, $A_\mathrm{lin}$ and $A_\mathrm{opt}$, used in the simulation. They are plotted as a function of normalized time $s$.
    (c)Plots of the infidelity $\delta_N$ when $A_\mathrm{lin}$ or $A_\mathrm{opt}$ is adopted.
    (d)The squared wavefunction of a electron after ATE over $N$ steps with $A_\mathrm{lin}$. 
    The black dashed line represents the exact ground state obtained from the numerical diagonalization of the final Hamiltonian. The blue circles represent the ground state of initial Hamiltonian.
    The orange, green, and red circles represent the output state of ATE over 1000, 5000, and 10000 steps, respectively.
    }
    \label{fig:results_harm}
\end{figure}

Under these conditions, $f(A)$ in Eq. (\ref{def_f}) and scheduling functions in Eq.(\ref{A_opt}) are numerically calculated and shown in Fig. \ref{fig:results_harm}(a) and (b), respectively. 
Since $f(A)$ is larger in regions where $A$ is smaller, $A_\mathrm{opt}$ is designed to change slowly at first.
The adiabatic conditions estimated from Eq. (\ref{ac_final}) are $t_\mathrm{f} \gg 92.01 \, $ for $A_\mathrm{lin}$ and $t_\mathrm{f} \gg 2.84 \,$ for $A_\mathrm{opt}$ (the schedules can be seen in Fig \ref{fig:results_harm}(b)).
In other words, it is estimated that about 30 times more time is required with $A_\mathrm{lin}$ than with $A_\mathrm{opt}$.
Fig. \ref{fig:results_harm}(c) shows the dependence of $\delta_N$ on the number of steps $N$ after performing ATE with these two scheduling functions.
$N$ required for $\delta_N$ to fall below $10^{-2}$ is about 9000 for $A_\mathrm{lin}$ and 300 for $A_\mathrm{opt}$, which is roughly consistent with the estimate of $t_\mathrm{f}$ based on the adiabatic condition of Eq. (\ref{ac_final}).
Fig. \ref{fig:results_harm}(d) shows the change in output state $| \psi (t_\mathrm{f}; N) \rangle$ as $N$ is increased.
Since the value of $\Delta t$ is fixed, a larger $N$ means a larger $t_\mathrm{f}$, i.e., the Hamiltonian $\hat{\mathcal{H}}(t)$ changes more slowly.
Starting from an initial state where the probability of existence of an electron is uniform, we can see that the output state approaches the exact ground state of $\hat{\mathcal{H}}(t_\mathrm{f})$ as $N$ increases.

$\delta_N$ is expected to decrease as $N$ and $t_f$ increase, but seems to reach a plateau and converges to around $3\times 10^{-4}$ in the region of $N > 10^4$ with $A_\mathrm{opt}$ in Fig. \ref{fig:results_harm}(c). 
This might be the result of errors caused by the fact that $\Delta t$ is now fixed to a finite value, that the right-hand side of Eq. (\ref{def_psi_tf}) is approximated by a finite product, and that the first-order Suzuki-Trotter expansion is used to implement the simulation circuit (see Appendix \ref{finite_delta_t}).

\subsection{Search for the optimal structure of an \texorpdfstring{${\rm H}_2^+$}{TEXT} molecule}
As an example of a simulation to find the most stable structure among many atomic configurations, we consider the optimization of the bond length of an H$_2^+$ molecule in one-dimensional space. The interactions between particles are modelled by the soft-Coulomb interaction: 
\begin{gather}
    V_\mathrm{soft} = \frac{Z_1 Z_2}{\sqrt{|\boldsymbol{r}_1-\boldsymbol{r}_2|^2 + \lambda_{12}^2}},
\end{gather}
where $\boldsymbol{r}_i$ and $Z_i$ are the position and the charge of the $i$-th particle, respectively. $\lambda_{ij}$ is the softness of the interaction and set as $\lambda_{ij} =1$ for all interactions in this simulation.
Thus, the potential $\hat{V}_{\mathrm{en},J}$ received by an electron confined in a one-dimensional space of length $L$ is expressed by the following equation where $0 \leq x <L$ and hydrogen nuclei are located at $(L+d_J)/2$ and  $(L-d_J)/2$:
\begin{gather}
    \hat{V}_{\mathrm{en},J} \equiv \frac{-1}{\sqrt{\left(\hat{x}-L/2-d_J/2\right)^2 + 1}} \nonumber \\
    + \frac{-1}{\sqrt{\left(\hat{x}-L/2+d_J/2 \right)^2 + 1}},
\end{gather}
where $d_J$ denotes the bond length of the H$_2^+$ molecule.
The following $\hat{\mathcal{H}}_J$ represents the Hamiltonian for a quantum mechanical single electron and classical two hydrogen nuclei system when the bond length is $d_J$.
\begin{gather}
    \hat{\mathcal{H}}_J = \hat{T}_\mathrm{el} + \hat{V}_{\mathrm{en},J} + \frac{1}{\sqrt{d_J^2 +1}}.
\end{gather}
In this simulation, we set $L=15$ and $n_{qe}=6$.
Now, considering four candidate bond lengths $(d_0, d_1, d_2, d_3) = (2.0, 4.0, 6.0, 8.0)$, and numerically diagonalizing $ \hat{\mathcal{H}}_J$ for each of them to get the ground state energy $E_J^\mathrm{gs}$, we obtain $(E_0^\mathrm{gs}, E_1^\mathrm{gs}, E_2^\mathrm{gs}, E_3^\mathrm{gs}) = (-0.811, -0.750, -0.693, -0.679)$, which means $J=0$ is corresponding to the optimal structure among them.

This structure search problem is solved by ATE based on the method described in Sec. \ref{method_str_opt}.
The number of qubits in the register to represent the nuclear configuration is set to $n_{qn}=2$, and the four configurations
$(d_0, d_1, d_2, d_3) = (2.0, 4.0, 6.0, 8.0)$ are assigned to the $|0\rangle$,  $|1\rangle$,  $|2\rangle$, and  $|3\rangle$ states of the nuclear register, respectively.
The operator $\hat{V}_\mathrm{nn}$ representing the interaction between nuclei can be expressed as
\begin{gather}
    \hat{V}_\mathrm{nn} = \sum_{J=0}^3 \frac{1}{\sqrt{d_J^2 +1}} |J \rangle \langle J|.
\end{gather}
The interaction $\hat{V}_\mathrm{en}$ between the nuclei and the electron can be written as
\begin{gather}
    \hat{V}_\mathrm{en} = \sum_{J=0}^3 \hat{V}_{\mathrm{en},J} \otimes | J \rangle \langle J |.
\end{gather}
With the operators defined above, the time-dependent Hamiltonian considered here is given by
\begin{gather}
    \hat{\mathcal{H}}(t)=\hat{T}_{\mathrm{el}} \otimes \hat{I}_{\mathrm{nucl}} 
    +A(t) \left[\hat{V}_{\mathrm{en}} +\hat{I}_{\mathrm{el}} \otimes \hat{V}_{\mathrm{nn}} \right] \nonumber \\
    -\left(1-A(t)\right) \hat{I}_{\mathrm{el}} \otimes J_x  \sum_{l=1}^{n_{qn}} \hat{X}_l.
\end{gather}
Thus, the initial Hamiltonian $\hat{\mathcal{H}}(0)$ and its ground state $|\psi(0) \rangle$ are given by
\begin{gather}
    \hat{\mathcal{H}}(0)=\hat{T}_{\mathrm{el}} \otimes \hat{I}_{\mathrm{nucl}} 
    - \hat{I}_{\mathrm{el}} \otimes J_x  \sum_{l=1}^{n_{qn}} \hat{X}_l, \\
    |\psi(0) \rangle = |\psi_0^\mathrm{gs} \rangle = |+ \rangle^{\otimes n_{qe}} \otimes |+ \rangle^{\otimes n_{qn}} .
\end{gather}
The initial ground state can be prepared using Hadamard gates:
\begin{gather}
    U_\mathrm{init} = H^{\otimes (n_{qe}+n_{qn})}.
\end{gather}

\begin{figure}[ht]
    \centering
    \includegraphics[width=0.45 \textwidth]{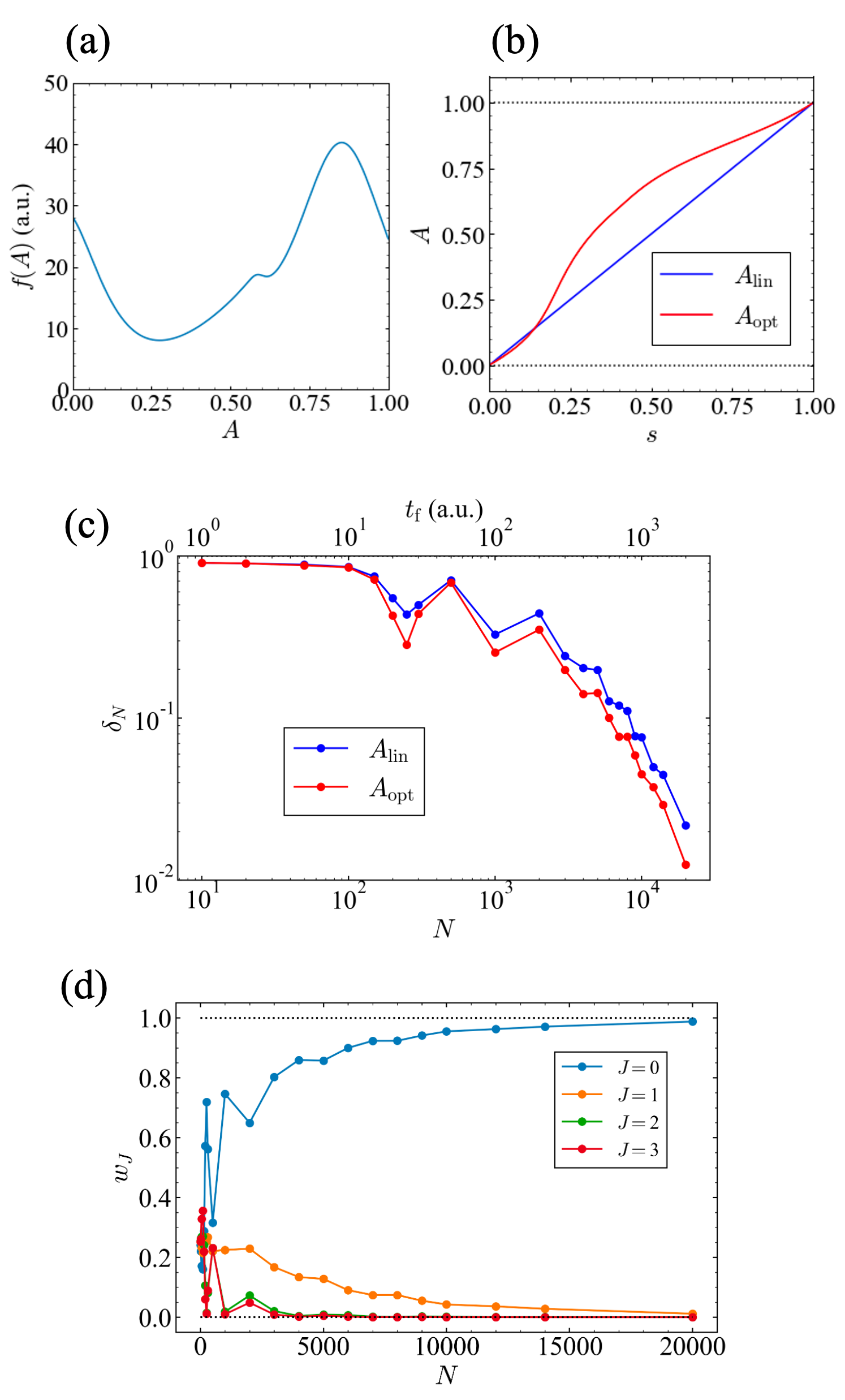}
    \caption{
    The simulation results of ATE for the search for the optimal bond length of an $\text{H}_2^+$ molecule.
    (a) The indicator of adiabatic condition $f(A)$ defined in Eq. (\ref{def_f}). The maximum value is 40.23 
    (b) The two scheduling functions, $A_\mathrm{lin}$ and $A_\mathrm{opt}$, used in the simulation. They are plotted as a function of normalized time $s$.
    (c) Plots of the infidelity $\delta_N$ when $A_\mathrm{lin}$ or $A_\mathrm{opt}$ is adopted.
    (d) Plots of the probability $w_J$ of obtaining the $J$-th structure when the nuclear register is observed with scheduling $A_\mathrm{lin}$.
    }
    \label{fig:results_H2opt}
\end{figure}

Under these conditions, the simulation is run with $\Delta t = 0.1$ and $J_x = 0.1$.
As in the case of Sec. \ref{sec_harm_ele}, we consider $f(A)$ defined in Eq. (\ref{def_f}) and the two scheduling functions in Eq. (\ref{A_opt}).
For this purpose, $|\psi_{st_\mathrm{f}}^\mathrm{gs}\rangle, |\psi_{st_\mathrm{f}}^{j}\rangle$, and $\Delta_j (st_\mathrm{f})$ are obtained by numerical diagonalization of $\hat{\mathcal{H}}(st_\mathrm{f})$, and $V$ is defined as
\begin{gather}
    V = \frac{\partial}{\partial A} \hat{\mathcal{H}}(t)= \hat{V}_{\mathrm{en}} +\hat{I}_{\mathrm{el}} \otimes \left[ \hat{V}_{\mathrm{nn}} + J_x  \sum_{l=1}^{n_{qn}} \hat{X}_l \right].
\end{gather}
$f(A)$ in Eq. (\ref{def_f}) and scheduling functions in Eq.(\ref{A_opt}) are numerically calculated and shown in Figs. \ref{fig:results_H2opt}(a) and (b), respectively. The adiabatic conditions estimated from Eq. (\ref{ac_final}) are $t_\mathrm{f} \gg 40.23 \,$ for $A_\mathrm{lin}$ and $t_\mathrm{f} \gg 20.40 \,$ for $A_\mathrm{opt}$, which means $A_\mathrm{lin}$ and $A_\mathrm{opt}$ require the similar order of magnitude for $t_\mathrm{f}$.
Fig. \ref{fig:results_H2opt}(c) shows the dependence of $\delta_N$ on the number of steps $N$ after the ATE performed with these two scheduling functions.
Actually, for the same $N$, $\delta_N$ is slightly smaller when $A_\mathrm{opt}$ is used.
For a more quantitative discussion, it may be necessary to consider the errors caused by first-order Suzuki-Trotter expansion and the approximation by finite product on the right-hand side of Eq. (\ref{def_psi_tf}).
It may also be useful to perform the analysis under more rigorous adiabatic conditions such as those found in Ref.~\cite{Jansen2007JMathPhys, Kimura2022JPhysA}, or to consider the sum of the effects of all transitions both from the ground state and between excited states, rather than just one dominant transition from the ground state like Eq. (\ref{adiabatic_condition}).

Importantly, this scheme based on ATE allows us to successfully search for the most stable structure. 
Fig. \ref{fig:results_H2opt}(d) shows the weights $w_J = \langle \psi (t_\mathrm{f}; N) | \hat{P}_J | \psi (t_\mathrm{f}; N) \rangle$
of the $J$-th structure in the output state when $A_\mathrm{lin}$ is adopted,
where $\hat{P}_J$ is a projection operator to the $J$-th structure and defined as $\hat{P}_J \equiv \hat{I}_\mathrm{el}\otimes |J\rangle \langle J|$.
This $w_J$ is equal to the probability of obtaining the $J$-th structure when the nuclear register is observed.
It is clear that when $N$ is sufficiently large, the most stable structure, $J=0$ has the highest probability.
\section{Conclusions}

This paper describes a construction of quantum circuit for the ground state of an electronic system using ATE based on the first quantization.
A method to create an initial ground state that is antisymmetric and non-degenerate is also proposed.
Furthermore, by considering the problem of total energy minimization for quantum-mechanical electrons and classical nuclei, we also construct a circuit to find the optimal configuration of nuclei.
In the simulations of simple systems, we successfully obtained the wavefunction of the ground state in the calculation of an electronic system, and demonstrated that the optimal structure can be obtained with high probability after a sufficiently long ATE in the calculation for the optimization of molecular bond length.
Although previous studies using PITE exist for similar problems~\cite{Kosugi2022PhysRevResearch, kosugi2023mag, kosugi2022exhaustive,nishi2022acceleration, nishi2023analyzing}, ATE can be another alternative of nonvariational ground-state calculation algorithms.
Since our scheme based on ATE does not require ancillary qubits due to its unitarity, unlike PITE, it has the advantage that it is only necessary to implement a simple RTE operator rather than CRTE operators.
If we apply the same method used to implement RTE in PITE based on first quantization~\cite{Kosugi2022PhysRevResearch, kosugi2022exhaustive}, to our ATE circuit, the number of operations in one ATE step and one PITE step has the same scaling with respect to the system size and the number of electrons $n_e$. It is typically estimated as $\mathcal{O}(n_e^2 \mathrm{poly}(\mathrm{log}n_e))$.
As a further study, it would be valuable to compare ground-state computation algorithms in terms of the overall circuit depth to obtain the output state, rather than focusing solely on one step of the computation.

\begin{acknowledgments}
The authors wish to thank Hidetoshi Nishimori for the insightful discussion.
This work was supported by MEXT under "Program for Promoting Researches on the Supercomputer Fugaku" (JPMXP1020200205) and by JSPS KAKENHI under Grant-in-Aid for Scientific Research (A) No. 21H04553. This study was carried out using the TSUBAME3.0 supercomputer at Tokyo Institute of Technology and the facilities of the Supercomputer Center, the Institute for Solid State Physics, the University of Tokyo.

\end{acknowledgments}

\appendix
\section{Proof of Prop. (\ref{injection})}
\label{sec:proof_of_LI}

\begin{proof}
\begin{gather}
\epsilon(n_x, n_y, n_z) = \epsilon(m_x, m_y, m_z) \nonumber \\
\Leftrightarrow (n_x-m_x)+\sqrt{2}(n_y-m_y) +\sqrt{3}(n_z-m_z) = 0 \nonumber \\
\Leftrightarrow k_x+\sqrt{2}k_y +\sqrt{3}k_z = 0,
\label{eq:k_mu}
\end{gather}
where $k_\mu \equiv n_\mu- m_\mu \in \mathbb{Z}$.
It is proved below that the only $(k_x, k_y, k_z)$ satisfying Eq. (\ref{eq:k_mu}) is $(0,0,0)$ proving our claim.
\begin{gather}
k_x+\sqrt{2}k_y +\sqrt{3}k_z = 0 \nonumber \\
\Rightarrow (\sqrt{2}k_y +\sqrt{3}k_z)^2 = k_x^2 \nonumber \\
\Rightarrow  \sqrt{6} k_yk_z= \frac{k_x^2-2k_y^2-3k_z^2 }{2}.
\label{eq:k_squared_condition}
\end{gather}
\\
If $k_yk_z \neq 0$, by Eq. (\ref{eq:k_squared_condition}), 
\begin{gather}
\sqrt{6} = \frac{k_x^2-2k_y^2-3k_z^2 }{2k_yk_z}.
\label{eq:kykz_nonzero}
\end{gather}
Since the left-hand side is irrational and the right-hand side is rational, no $(k_x, k_y, k_z)$ satisfies Eq. (\ref{eq:kykz_nonzero}).
\\
\\
If $k_y=0$ and $k_z\neq 0$, by Eq. (\ref{eq:k_mu})
\begin{gather}
\sqrt{3}=-k_x / k_z .
\label{eq:ky_zero_kz_nonzero}
\end{gather}
Since the left-hand side is irrational and the right-hand side is rational, no $(k_x, k_y, k_z)$ satisfies Eq. (\ref{eq:ky_zero_kz_nonzero}).
\\
\\
If $k_y\neq0$ and $k_z= 0$, by Eq. (\ref{eq:k_mu})
\begin{gather}
\sqrt{2}=-k_x / k_y .
\label{eq:ky_nonzero_kz_zero}
\end{gather}
Since the left-hand side is irrational and the right-hand side is rational, no $(k_x, k_y, k_z)$ satisfies Eq. (\ref{eq:ky_nonzero_kz_zero}).
\\
\\
If $k_y=0$ and $k_z= 0$, $k_x=0$ by Eq. (\ref{eq:k_mu}).
\\
\\
The above proves
\begin{gather}
k_x+\sqrt{2}k_y +\sqrt{3}k_z = 0 \Rightarrow (k_x, k_y, k_z)=(0,0,0),
\end{gather}
and thus our claim.
\end{proof}

\section{Effect of finite \texorpdfstring{$\Delta t$}{TEXT} on infidelity}
\label{finite_delta_t}
\begin{figure}[ht]
    \centering
    \includegraphics[width=0.45 \textwidth]{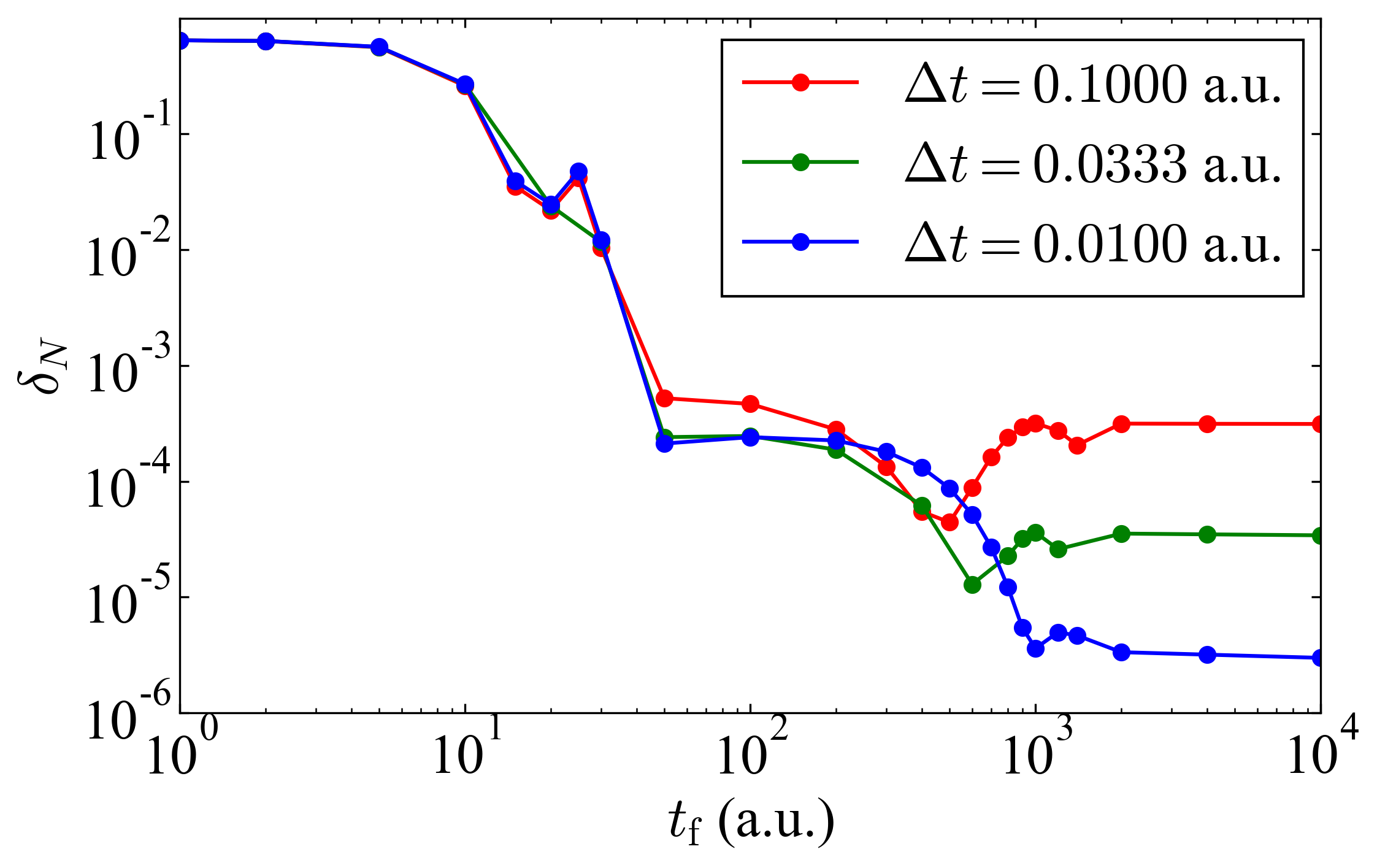}
    \caption{Plots of the infidelity $\delta_N$ with $A_\mathrm{opt}$ and different values of $\Delta t$}
    \label{fig:delta_t_dependence}
\end{figure}

We examine the behavior of the infidelity $\delta_N$ with the scheduling function $A_\mathrm{opt}$ at different values of $\Delta t$ for one electron model in a parabolic potential described in Sec. \ref{sec_harm_ele}.
In this simulation, the unitary time evolution has $\mathcal{O}(\Delta t^2)$ error because it is approximated in the form of a finite product as in Eq. (\ref{def_psi_tf}) and the first-order Suzuki-Trotter expansion is used in the circuit implementation of $\exp (-i \hat{\mathcal{H}}(t_m) \Delta t)$.
Fig. \ref{fig:delta_t_dependence} shows that the smaller $\Delta t$ is, the smaller $\delta_N$ can be realized in the region of $t_\mathrm{f}>10^3$.
This result suggests that the saturation of $\delta_N$ is due to $\mathcal{O}(\Delta t^2)$ terms.
Of course, ideally $\Delta t$ should be as small as possible, but it has to be noted that the total circuit depth required to achieve the same $t_\mathrm{f}$ is proportional to $\Delta t^{-1}$ as the number of ATE steps $N$ is equal to $\lceil t_\mathrm{f}/\Delta t \rceil$.

\bibliographystyle{apsrev4-2}
\bibliography{ref}

\end{document}